\documentclass[prl,twocolumn,letterpaper,showpacs,superscriptaddress]{revtex4}
\def\be{\begin{equation}}
\def\fin{\end{equation}}
\def\disp{\displaystyle}

\def\T{{\sf T\kern-.45em T}}
\def\C{\kern.1em{\raise.47ex\hbox{$\scriptscriptstyle |$}}
             \kern-.40em{\sf C}}

\def\ze{\zeta}
\def\al{\alpha}
\def\b{\beta}

\def\hfl{\disp\mathop{\hbox to 10mm{\rightarrowfill}}}

\usepackage{graphicx,amssymb,amsmath}
\begin{document}
\title{Reversible Diffusion-Limited Reactions:
"Chemical Equilibrium" State and the Law of Mass Action Revisited}

\date{\today}

\author{R.Voituriez}
%\email{voiturie@lptl.jussieu.fr}
\affiliation{Laboratoire de Physique Th{\'e}orique des Liquides,
Universit{\'e} Paris 6, 4 Place Jussieu, 75252 Paris, France}
\author{M.Moreau}
%\email{moreau@lptl.jussieu.fr}
\affiliation{Laboratoire de Physique Th{\'e}orique des Liquides,
Universit{\'e} Paris 6, 4 Place Jussieu, 75252 Paris, France}
\author{G.Oshanin}
%\email{oshanin@lptl.jussieu.fr}
\affiliation{Laboratoire de Physique Th{\'e}orique des Liquides,
Universit{\'e} Paris 6, 4 Place Jussieu, 75252 Paris, France}
\affiliation{
Max-Planck-Institut f\"ur Metallforschung, Heisenbergstr. 3,
D-70569 Stuttgart, Germany}
\affiliation{Institut f\"ur Theoretische und Angewandte Physik,
Universit\"at Stuttgart, Pfaffenwaldring 57, D-70569 Stuttgart,
Germany}

\begin{abstract}
The validity of two fundamental concepts of classical chemical
kinetics - the notion of "Chemical Equilibrium" and the "Law of
Mass Action" - are re-examined for reversible
\textit{diffusion-limited} reactions (DLR), as exemplified here by
association/dissociation $A+A \rightleftharpoons B$ reactions. We
consider a general model of long-ranged reactions, such that any
pair of $A$ particles, separated by distance $\mu$, may react with
probability $\omega_+(\mu)$, and any $B$ may dissociate with
probability $\omega_-(\lambda)$ into a geminate pair of $A$s
separated by distance $\lambda$. Within an exact analytical
approach, we show that
 the asymptotic state attained by reversible
DLR at $t = \infty$ is generally \textit{not a true
thermodynamic equilibrium}, but rather a non-equilibrium
steady-state, and that the Law of Mass Action is invalid. The
classical picture holds \text{only} in physically unrealistic case
when $\omega_+(\mu) \equiv \omega_-(\mu)$ for any value of $\mu$.
\end{abstract}


\maketitle

"Chemical Equilibrium" (CE) and the "Law of Mass Action" (LMA) are
two central concepts of classical chemical kinetics (see, e.g.,
Ref.\cite{keizer}). In virtually every text-book one finds,
regarding, e.g., the behavior of reversible
association/dissociation reaction of the form \be\label{chem} A+A\
\disp \mathop{\rightleftharpoons}^{K_+}_{K_-} \ B, \fin where
$K_+$ and $K_-$ are the forward and the backward rate constants,
respectively, that the state achieved at $t = \infty$ is the CE.
Thermodynamically, the CE is the state in which the net Gibbs free
energy change of the $A$ and $B$
 mixture vanishes.
Kinetically,
the CE is the state with no net change in activity,
in which the rates of the forward, $K_+ a_{\infty}^2$, and the backward, $K_-
b_{\infty}$, reactions are equal and opposite, such that
$a_{\infty}$ and $b_{\infty}$ - the $A$ and $B$
 "equilibrium" concentrations, obey the LMA:
\be
\label{lma}
-K_+ a_{\infty}^2 + K_- b_{\infty} = 0 \;\;\; \text{or} \;\;\;
\frac{a_{\infty}^2}{b_{\infty}} = \frac{K_-}{K_+} = K_{eq},
\fin
with $K_{eq}$ being the  "equilibrium" constant,
dependent on the \textit{thermodynamic}
properties only \cite{keizer}.\\

In this paper we re-examine the validity of these two fundamental
concepts for reversible diffusion-limited reactions (DLR)
\cite{bamberg}. Our objective is to determine whether for
reversible DLR the CE is always a true thermodynamic equilibrium
state (TES) and whether the LMA in Eq.(\ref{lma}) always holds. We
concentrate on a simple reaction scheme in Eq.(\ref{chem}), but
our analysis can be readily generalized (and our conclusions will
remain valid) for any other type of reversible DLR. We consider a
general reaction model with long-ranged reaction probabilities:
That is, any pair of randomly moving $A$s, separated by distance
$\mu$, may associate with probability $\omega_+(\mu)$, while any
product molecule $B$ may dissociate spontaneously producing, with
probability $\omega_-(\lambda)$, a geminate pair of $A$s separated
by distance $\lambda$. For this model, in terms of a formally
exact approach, we deduce a criterion determining
 when the classical LMA in Eq.(\ref{lma}) holds and
when the asymptotic $t = \infty$ state is a true TES. We show that
this is only the case when $ \omega_+(\mu) \equiv \omega_-(\mu)$
for any $\mu$, which is apparently a non-realistic condition. On
contrary, when the bimolecular forward and the unimolecular
backward reaction probabilities do not coincide exactly, (which is
more appropriate on physical grounds), we find that the LMA in
Eq.(\ref{lma}) is violated and that the CE is not a true TES but
rather a nonequilibrium steady-state. This finding is, of course,
in a striking contradiction with the generally accepted classical
picture.

The classical picture
was, however, already shown to be inadequate
in many situations.
For example, for reversible reactions
it predicts an exponential
approach toward the CE state.
It was realized
that this is not the case for
reversible DLR; here,
the concentrations approach
the asymptotic $t = \infty$ state
only
 as a power law ($t^{-d/2}$ in $d$ dimensions) \cite{ovch1,tachya,redner,gleb,szabo,shin,naumann,gopich,cardy,gopich2,gopich3}. This \textit{anomalous}
behavior, which
stems out of
many particle and non-linear effects,
was
indeed observed
in excited state proton transfer reactions \cite{1}.

For the reversible DLR the validity of LMA in Eq.(\ref{lma})  was
questioned in Refs.\cite{ovch1} and \cite{gleb}, where some
non-vanishing corrections were obtained within
\textit{approximate} approaches. On contrary, \textit{exact
solutions} obtained in Refs.\cite{cardy} and \cite{gopich3} in
\textit{contact reaction} approximation,
 have shown that the LMA holds
 and that the CE state is a true TES.
Note, however, that \textit{contact reaction} approximation is
just a mathematical trick employed to obtain a tractable formalism
at expense of a large degree of arbitrariness especially regarding
the choice of the "reaction radius". In reality, an elementary
reaction act results from an interplay of many factors and is
influenced by solvent structure, potential interactions, a variety
of particles' energies and angular orientations, quantum processes
of different origin and etc \cite{hynes}, such that the reaction
constants are actually long-ranged and the very notion of a fixed
"reaction radius" does not make much sense. To elucidate such a
delicate issue as the nature of the asymptotic $t = \infty$ state,
(and to reconcile the discord between Refs.\cite{ovch1,gleb} and
\cite{cardy,gopich3}), one has to consider the realistic
distance-dependent reaction probabilities, which has not been done
so far within a rigorous approach.

Consider
a $d$-dimensional hypercubic lattice of spacing $\ell$,
containing $A$ and $B$ particles, which perform unconstrained
random walks between neighboring sites.
At any time moment $t$,
any two $A$s
may react
with probability $\omega_+(\mu)$,
where $\mu$ is an instantaneous distance between these $A$s,
and
 form a $B$ particle,
placed at the half-distance between two $A$s. Any $B$, in turn,
may dissociate with probability $\omega_-(\lambda)$ producing a
randomly oriented pair of $A$s separated by distance $\lambda$.
The long-ranged reaction constants are: $k_{+}(\mu) = k_{+}
\omega_{+}(\mu)$ and $k_{-}(\lambda) = k_{-} \omega_{-}(\lambda)$,
where $k_{+}$ and $k_{-}$ are some amplitudes. The overall
constants are thus given by $K_{+} = \sum_{\mu} k_{+}(\mu)$ and
$K_{-} = \sum_{\lambda} k_{-}(\lambda)$. We suppose that
$k_{+}(\mu)$ and $k_{-}(\lambda)$ are \textit{arbitrary} summable
functions of $\mu$ and $\lambda$.

The state of the system is determined by time- and site-dependent
occupation numbers $A(x)$  and $B(x)$. Let $P_t[A,B]$ denote the
probability of finding the system in configuration $A=\{A(x)\}$
and $B=\{B(x)\}$ at time moment $t$. The time evolution of
$P_t[A,B]$ due to reactions and diffusion processes
 on the microscopic,
many particle level is governed by the following master equation:
\begin{widetext}
\begin{eqnarray}
\dot{P}_t[A,B] &=& \frac{D}{\ell^2}\sum_x\sum_{e_x}
    \Big[\Big(A(e_x)+1\Big) P_t[A(x)-1,A(e_x)+1,B] + \Big(B(e_x)+1\Big) P_t[A,B(x)-1,B(e_x)+1] \nonumber \\
&-&  \Big(A(x)+B(x)\Big) P_t[A,B] \Big]
+  \sum_\mu k_+(\mu)\sum_x
    \Big[ \Big(A(x-\frac{\mu}{2})+1\Big)
    \Big(A(x+\frac{\mu}{2})+1\Big) \times \nonumber\\
 &\times& P_t[A(x-\frac{\mu}{2})+1,A(x+\frac{\mu}{2})+1,B(x)-1]
 -  A(x-\frac{\mu}{2})A(x+\frac{\mu}{2})P_t[A,B] \Big] \nonumber\\
&+&  \sum_{\lambda} k_{-}(\lambda)\sum_x \Big[\Big(B(x)+1\Big)
P_t[A(x-\frac{\mu}{2})-1,A(x+\frac{\mu}{2})-1,B(x)+1] -
B(x)P_t[A,B]\Big] \label{ME2}
\end{eqnarray}
\end{widetext}
where $D = \ell^2/2 d \tau$ is $A$ and $B$ particles diffusion
coefficient, $\tau$ is a hopping time,  while $\sum_{e_x}$ denotes
summation over orientations of the lattice vector  $e_x$.

Our analytical approach to the solution of Eq.(\ref{ME2}) is based
on formally exact Poisson representation method, proposed
originally in Ref.\cite{gardiner} for systems without diffusion,
and elegantly generalized in Ref.\cite{gopich3} for reversible DLR
in contact approximation. Extension of this approach to
long-ranged reversible DLR is straightforward and here we merely
outline the steps involved.

Our next step consists in
projecting
$A(x)$ and $B(x)$
onto the Poisson states $\alpha = \{\alpha(x)\}$ and $\beta = \{\beta(x)\}$ \cite{gardiner}:
\begin{eqnarray}
P_t[A,B] &=& \int \prod_x d \alpha(x) d\beta(x) \frac{\exp\left(-\alpha(x)\right) \alpha(x)^{A(x)}
}{A(x)!} \times \nonumber\\
&\times& \frac{\exp\left(-\beta(x)\right) \beta(x)^{B(x)}
}{B(x)!} F_t[\alpha,\beta],
\end{eqnarray}
which results, after some straightforward calculations, in the
following Fokker-Planck equation for $F_t[\al,\b]$:
\begin{eqnarray}\label{FP1}
\dot{F} &=& \disp \sum_{x} \Big\{
\frac{1}{2} \frac{\partial^2 \Big(C_2 F\Big)}{\partial\al(x-\disp \frac{\mu}{2})\partial\al(x+\disp \frac{\mu}{2})} - \nonumber\\
&-&\frac{\partial \Big(C_1 F\Big)}{\partial\al(x)} - \frac{\partial \Big(C_2 F\Big)}{\partial\b(x)} +\nonumber\\
+& \disp\frac{D}{\ell^2}& \Big[ \frac{\partial}{\partial \al(x)}
\Big(\Delta \al(x)  F \Big) + \frac{\partial}{\partial \b(x)}
\Big(\Delta \b(x) F\Big) \Big]\Big\},
\end{eqnarray}
where $\Delta$ is the second finite difference operator, and
\begin{eqnarray}
C_1 &=& -\disp \sum_{\mu} k_{+}(\mu)\al(x)\Big(\al(x-\mu)+\al(x+\mu)\Big) + \nonumber\\
           &+& \disp \sum_{\mu} k_{-}(\mu)\Big(\b(x-\mu)+\b(x+\mu)\Big),\\
C_2 &=& \disp \sum_{\mu} \Big(k_{+}(\mu)\al(x+\disp
\frac{\mu}{2})\al(x-\disp \frac{\mu}{2})-k_{-}(\mu)\b(x)\Big)
\end{eqnarray}
Using It\^o's equivalence, we find next the following non-linear
Langevin equations corresponding to Eq.(\ref{FP1}): \be\label{lan}
\left\{\begin{array}{l}
\dot{\al}(x)-D\Delta \al(x)=  \disp C_1 + \ze(x,t),\\
\disp \dot{\b}(x) - D \Delta \b(x)=  \disp C_2,
\end{array}
\right.
\fin
where  $\ze(x,t)$
is a Gaussian noise with zero mean and
\begin{eqnarray} \label{noises}
&&\Big<\ze(x,t)\ze(x+x',t')\Big>= \delta(t-t')\Big[
k_{-}(x')\Big<\b(x')\Big> \nonumber\\
&-&k_{+}(x')\Big<\al(x-x'/2)\al(x+x'/2)\Big>
\Big].
\end{eqnarray}
Taking into account that $<\alpha(x)> = a_t$ and $<\beta(x)> =
b_t$ \cite{gardiner}, where $a_t$ and $b_t$ are $A$ and $B$
particles' mean concentrations, we represent the Poisson fields as
the sum of mean values and fluctuations, such that
\begin{equation} \label{dev}
\begin{array}{l}
\alpha(x)=a_t+\delta\al(x,t),\\
\b(x)=b_t+\delta\b(x,t)
\end{array}\fin
Substituting next these expressions into Langevin Eqs.(\ref{lan})
and averaging, we find that $a_t$ and $b_t$ follow:
\begin{eqnarray} \label{exa}
\dot{a}_t = - 2 K_+ a^2_t + 2 K_- b_t + 2 \Omega_t(p=0), \;\;\;
\dot{b}_t = - 2 \dot{a}_t,
\end{eqnarray}
where
 \be \label{Omega} \Omega_t(p) \equiv - \disp \int d\mu e^{i
(\mu \cdot p) }k_{+}(\mu)\sigma_{\alpha \alpha}(\mu,t), \fin
and $\sigma_{\alpha \alpha}(\mu,t)$ is the pairwise correlation
function: \be\label{cor} \sigma_{\alpha \alpha}(\mu,t) = \Big<
\delta\alpha(x-\mu/2,t) \delta\alpha(x + \mu/2,t) \Big>. \fin
Equations (\ref{exa}) are formally exact for any $t$ and show that
the time evolution of observables $a_t$ and $b_t$ is coupled to
that of correlations. Fot $t = \infty$, we get from
Eqs.(\ref{exa}): \be\label{eps} \disp
\frac{a_{\infty}^2}{b_{\infty}} = \frac{K_{-}}{K_{+}} +
\frac{\Omega_{\infty}(0)}{K_{+} b_{\infty}}, \fin  which resembles
the classical LMA in Eq.(\ref{lma}), but differs from it due to
the term ${\Omega_{\infty}(0)}/K_{+} b_{\infty}$, which
embodies all non-trivial physics associated with fluctuation
effects; the classical LMA holds
if and only if  $\sigma_{\al
\al}(\mu,\infty) \equiv 0$.

Hence, we focus on $\sigma_{\alpha \alpha}(\mu,\infty)$. From
Eqs.(\ref{lan}) we have that the Fourier-transformed
$\sigma_{\alpha \alpha}(\mu,t)$ follows
\begin{eqnarray}
\label{sig}
&&\sigma_{\alpha \alpha}(p,t) = \Big<\Big[G_{\al\b}(p,t)*\Big(C_2(p)+\delta\b_0(p)\delta(t)\Big)\Big]^2\Big>\nonumber\\
&+&\Big<\Big[G_{\al\al}(p,t)*\Big(C_1(p)+\zeta(p,t)+
\delta\al_0(p)\delta(t)\Big)\Big]^2\Big> \nonumber\\
&+&2\Big<\Big[G_{\al\al}(p,t)*\Big(C_1(p)+\zeta(p,t)+
\delta\al_0(p)\delta(t)\Big)\Big]\nonumber\\
 & \times&\Big[G_{\al\b}(p,t)*\Big(C_2(p)+\delta\b_0(p)\delta(t)\Big)\Big]\Big>,
\end{eqnarray}
where $"*"$ denotes the time convolution,
while $\delta\al_0$ and $ \delta\b_0$
stand for the initial values
of the
Fourier-transformed fluctuations.
Note, however, that the latter
produce
exponentially decreasing with $t$ terms
and hence, are insignificant.
In turn, the propagator $G(p,t)$
is an inverse of the matrix $M(p,t)$,
 $G(p,t)=M^{-1}(p,t)$, which is defined
in the Fourrier-Laplace space as:
\be\label{m}
\left(\begin{array}{ll}
s+Dp^2+2 a_\infty(K_{+}+k_+(p)) & - 2 k_-(p)\\
- 2 a_\infty k_+(p)& s+Dp^2+K_-
\end{array}\right),
\fin
$k_{\pm}(p)$ being the Fourier-transformed reaction constants.

We proceed further with a diagrammatic expansion of Eq.(15),
which has been previously developed in Ref.\cite{cardy}
for \textit{contact} DLR, and represents an expansion  with
respect to deviations from the equilibrium situation in
Eq.(\ref{lma}). In our case of long-range DLR, it can be deemed as
an expansion in powers of a small parameter: \be
\label{eeps} \epsilon(p) = k_{+}(p) a_{\infty}^2 - k_{-}(p)
b_{\infty}. \fin In doing so, we obtain, in the linear
order in $\epsilon(p)$, the following integral equation
for $\sigma_{\alpha
\alpha}(p,\infty)$: \be \label{+} \sigma_{\alpha
\alpha}(p,\infty)=\disp \Big[\Omega_{\infty}(p)-\epsilon(p)\Big] \int_0^{\infty} dt
G_{\al\al}^0(p,t)^2,
 \fin
where $G^0_{\al\al}(p,t)$ is the corresponding
propagator:
\begin{eqnarray}\label{0prop}
 G^0_{\al\al}(p,t)&=&\disp \sum_{\gamma=\pm} \frac{(K_--q^{\gamma})}{q^+-q^-}\disp e^{\disp -(Dp^2+q^{\gamma})t},
\end{eqnarray}
with
\begin{eqnarray}\label{q+}
q^\pm&=&\disp \Big[K_-+2a_\infty K_+\Big(1+w_+(p)\Big)\pm\sqrt{q}\Big]/2,\\
q&=&\disp \Big[K_-+2a_\infty K_+\Big(1+w_+(p)\Big)\Big]^2-\nonumber\\
&-&16K_+K_-a_\infty\Big(1-w_+(p)\Big)\Big(\frac{1}{2}+w_+(p)\Big).
\end{eqnarray}
Now we are in position  to deduce a criterion showing when the LMA
in Eq.(\ref{lma}) is violated already in the linear order. To do
this, we have to define from Eq.(\ref{+}) conditions when
$\sigma_{\alpha \alpha}(p,\infty) \neq 0$. Suppose, on contrary,
that $\sigma_{\alpha \alpha}(p,\infty) \equiv 0$. This implies, in
virtue of  Eq.(\ref{Omega}), that $\Omega_{\infty}(p) \equiv 0$.
Hence, in order to have $\sigma_{\alpha \alpha}(p,\infty) \equiv
0$, the parameter $\epsilon(p)$ in Eq.(\ref{eeps}) should be equal
to zero. i.e., \be \label{cond1} k_{-}(p) b_{\infty} \equiv
k_{+}(p) a^2_{\infty} \;\;\; \text{for any $p$}, \fin or, in the
$\mu$-domain, the equality $k_{-}(\mu) b_{\infty} \equiv
k_{+}(\mu) a^2_{\infty}$ should hold for any $\mu$. One infers
then that the identity in Eq.(\ref{cond1}) may hold only if the
reaction probabilities obey: $\omega_{-}(\mu) \equiv
\omega_{+}(\mu) = \omega(\mu)$ for any $\mu$, i.e. the elementary
reactions are \textit{microscopically homogeneous}. We note that
Refs.\cite{cardy} and \cite{gopich3}, which predicted that the
classical LMA holds and that the CE is a true TES, focused
precisely on the case of \textit{microscopically homogeneous}
contact reactions with $\omega(\mu) = \delta(\mu)$. If, on
contrary, the elementary reactions are \textit{microscopically
inhomogeneous}, i.e., $\omega_{-}(\mu) \neq \omega_{+}(\mu)$, pair
correlations $\sigma_{\alpha \alpha}(\mu,\infty) \neq 0$, and
hence, the LMA in Eq.(\ref{lma}) is violated. Therefore, general
conclusions of Refs.\cite{ovch1} and \cite{gleb}, which analyzed
\textit{microscopically inhomogeneous} reactions using approximate
approaches, are also qualitatively correct.

Note now that, remarkably, the violation of the
LMA implies
that the corresponding "Chemical Equilibrium"
is not a true TES.
This can be readily understood if one notices
that already in the linear order $G(p,t)$ is
dependent on particles diffusion
coefficient $D$, (see Eq.(\ref{0prop})).
This implies that
for \textit{microscopically inhomogeneous} reactions,
$\sigma_{\alpha \alpha}(\mu,\infty)$ does
depend on such "kinetic" parameter as $D$, which means, in turn,
that the CE is not a true TES but rather a non-equilibrium steady-state.

To illustrate our general conclusions
we focus now on 3D systems with
exponential $k_{+}(\mu)$ and $k_{-}(\lambda)$:
\be
\label{long} k_{+}(\mu)=\disp \frac{K_+}{8\pi R^3}e^{\disp
-|\mu|/R}\ \ {\rm and} \ \ k_{-}(\lambda)=\disp \frac{K_-}{8\pi
\Lambda^3}e^{-\disp |\lambda|/\Lambda}. \fin
In this case, we find from Eq.(\ref{+}) that $a_{\infty}$
and $b_{\infty}$ obey:
\be \label{ss} \frac{a_{\infty}^2}{b_{\infty}} =
\frac{K_-}{K_+} \Big[ 1 + \frac{K_+}{16 \pi D R} \left(1 -
\frac{R}{\Lambda}\right) + {\cal O}\Big((\Lambda - R)^2\Big)\Big], \fin
which
holds for sufficiently small $R$ and $\Lambda$. Eq.(\ref{ss})
also signifies
that $a_{\infty}$
and $b_{\infty}$ depend
on $D$, which dependence fades out
when either $D \to \infty$ or $R = \Lambda$.
As well, we find that
the large-$\mu$
behavior of pair correlations follows:
\be \label{pr}
\sigma_{\alpha \alpha}(\mu,\infty) \approx
\disp  \frac{16 b_\infty a_{\infty}^2 K_-K^2_{+}(\Lambda^2-R^2)}{4\pi D^2 \Big(4 K_{+} a_{\infty} + K_{-}\Big)} \disp \frac{e^{\disp -\mu/\Lambda_{corr}}}{\mu},
\fin
where the correlation length $\Lambda_{corr}$ is also $D$-dependent:
\be
\Lambda_{corr} = \sqrt{\frac{D}{4 K_{+} a_{\infty} + K_{-}}}.
\fin
Note that $\Lambda_{corr}$ may be much greater than
$R$, Eq.(\ref{long}), when $\tau_{chem} =
(4 K_{+} a_{\infty} + K_{-})^{-1} \gg \tau_{diff} = R^2/D$,
which is a fingerprint of an essentially
cooperative
behavior.

In conclusion,
we re-examined the validity of two fundamental
concepts of classical chemical kinetics
- the
notion of "Chemical Equilibrium" and the "Law of Mass Action" -
on example of diffusion-limited reversible
$A+A \rightleftharpoons B$ reactions
with general, distance-dependent reaction probabilities.
In terms of a formally exact
approach based on Gardiner's Poisson representation method
\cite{gardiner}, we deduced a criterion
determining the conditions
 when the classical LMA holds and
when
the CE is a true TES.
We realized that this is the case only
when the elementary
reaction probabilities obey the condition
of \textit{microscopic homogeneity}:
$ \omega_+(\mu) \equiv
\omega_-(\mu)$ for any $\mu$, which is apparently
unrealistic since the \textit{bimolecular} forward
and \textit{unimolecular}
backward reactions
are supported by completely different
physical processes of classical and quantum origin.
On contrary, we found that for
\textit{microscopically inhomogeneous} reactions,
when $ \omega_+(\mu) \neq
\omega_-(\mu)$, the classical LMA
is violated and
the CE is not a true TES but rather
a nonequilibrium
steady-state.

Consequently, for reversible DLR
the diffusional relaxation of the system is not fast enough to offset
the perturbative effect of ongoing \textit{microscopically inhomogeneous}
elementary reactions even in the asymptotic $t = \infty$ state.
We emphasize that such a non-equilibrium
steady-state is observed  for a \textit{closed} system with the \textit{conserved} overall
concentration of particles, without any external
inflow of particles.
We also note that contrary to the
dynamical behavior
of reversible DLR, (for which an anomalous power-law
decay emerges
if some \textit{conservation laws} are present \cite{ovch1,tachya,redner,gleb,szabo,shin,naumann,gopich,cardy,gopich2,gopich3}), for \textit{microscopically inhomogeneous} reactions
the LMA would be evidently violated
and the
CE would be a non-equilibrium steady-state
also in \textit{absence} of conserved parameters.
We finally remark that
the DLR furnish a remarkable example of
systems, for which
an arbitrarily small but finite difference (of classical or quantum origin)
between the microscopic rates $ \omega_+(\mu)$
and $\omega_-(\mu)$
results in a fundamental change in the macroscopic
asymptotic $t = \infty$ behavior.

The authors gratefully acknowledge helpful discussions with
S.F.Burlatsky, S.Bratos, I.V.Gopich, A.Lesne and M.Tachiya. GO
thanks the AvH Foundation for the financial
support via the Bessel Research Award.

\end{document}